# Software Stage-Effort Estimation Using Association Rule Mining and Fuzzy Set Theory

Mohammad Azzeh
Department of Computing
University of Bradford
Bradford BD7 1DP, U.K.
M.Y.A.Azzeh@brad.ac.uk

Daniel Neagu
Department of Computing
University of Bradford
Bradford BD7 1DP, U.K.
D.Neagu@brad.ac.uk

Peter Cowling
Department of Computing
University of Bradford
Bradford BD7 1DP, U.K.
P.I.Cowling@brad.ac.uk

## ABSTRACT
Relaying on early effort estimation to predict the required number of resources is not often sufficient, and could lead to under or over estimation. Software Project managers may not recognize that software development process should be refined regularly and that software prediction made at early stage of software development is yet kind of guesses. Even good predictions are not sufficient with inherent uncertainty and risks. The stage-effort estimation allows project manager to re-allocate correct number of resources, re-schedule project and control project progress to finish on time and within budget. In this paper we propose an approach to utilize prior effort records to predict stage effort. The proposed model combines concepts of Fuzzy set theory and association rule mining. The results were good in terms of prediction accuracy and have potential to deliver good stage-effort estimation.

## Categories and Subject Descriptors

D.2.9 [Software Engineering]: Management—cost estimation.

## 1. INTRODUCTION
Software effort estimation has long been and still a complex task for software industries [6, 7, 8, 9]. Due to dramatic changes in software development tools, methods, and methodologies, software applications become more complex, time to market is shortened, and the need to produce software at reasonable cost with high quality is the target of most organizations [1, 2]. Consequentially, a reliable and accurate early software effort estimation model is required in inception phase and particularly when bidding for a contract or making appropriate decisions. Unfortunately, this is not enough. Several surveys and reports [3, 5] revealed that 16% of 8000 complete projects only were delivered within budget and time, while 31% were cancelled before completion, and 53% were overrun in budget and schedule which resulted in project failure. Other authors [3, 4] showed that some 60% of large projects significantly overrun their estimates (with an error percentage that can vary from 100% to 200%) and 15% of the software projects are never completed due to the gross misestimating of development effort [4].

Project managers and software developers often recognize that estimate made at the beginning of software development is quite sufficient to be relied on until the end of software project [11]. However, in most cases this is not true because software development is a process of gradual refinement [10, 11]. Even good early estimates are only guesses, with inherent uncertainty and risks. In other words, the developers cannot depend on these estimates throughout software development without any sense of update to current project progress. This eventually will lead to expected overestimation or underestimation problems. It is acknowledged that under-estimation [4] causes understaffing and consequentially takes longer to deliver project than necessary. For instance, if you provide a project more resources than it really needs without sufficient scope controls of how to use them, the project is then likely to cost more than it should [4]. On the other hand, overestimation could lead to miss opportunities to funds in other projects in the same company [7]. This presses the need to dynamically predict the software effort during project progress in order to update the project schedule, thus, to finish on time and within budget. In this context, we understand that software effort estimation is a dynamic process and it needs gradual refinement during software development process to keep the project schedule under control and reduce associated risks.

In this paper we investigate the significant impact of using effort records of prior stages to develop an evolving picture of the potential effort for next stage. Stage is often a software life cycle phase such as requirements, design, coding, testing, and implementation [12]. Recently, most of the proposed approaches used statistical methods to map between prior stages and next stage [13]. These methods revealed that is difficult to predict next stage effort based on prior stages effort; in addition they usually require a large amount of data or data that have a certain statistical distribution [12].

The objective of the present paper is to propose a model that can predict stage-effort based on prior stage efforts. We combine the concepts of Fuzzy set theory [25] and association rule mining [22] to build such a model. Using association rule mining allows us to explore the hidden knowledge between prior effort stages and next effort stage. The Fuzzy set theory [25] was used to deal with linguistic terms that derived after partitioning a dataset to a number of intervals. Each interval is represented by a corresponding Fuzzy set which will be used for approximate reasoning to predict effort of target stage. The rest of the paper is structured as follows: in section 2 we review the stage effort estimation approaches. Section 3 presents an overview of Fuzzy set theory. We then introduce an overview of association rule mining in section 4. The proposed approach is discussed in section 5. The results of the empirical validation are discussed in sections 6 and 7, followed by conclusions of our study and recommendations for future work.

## 2. RELATED WORK

Although much researches have been carried out in the context of software effort estimation [15, 16, 17, 18], very little research effort has been put into the area of software stage-effort estimation. The term stage has been used in different contexts where some of them considered it as phase of software development lifecycle [13, 14] and others considered as calendar month [12].

MacDonell [13] investigated the potential of using prior effort data records to develop stage effort estimation. His model was built over sixteen projects collected from a single organization. The developed model revealed that there was no improvement on estimation accuracy when using only regression techniques. In contrast, he showed that prediction could be improved when combining regression technique with expert estimates. Ohlsson et al. [14] used phase-based data (proxy) such as number of requirements, flowcharts, number of test case, etc. to build a stage-effort prediction model using regression analysis. They injected each stage with several related proxies to improve prediction accuracy. The analysis based on 26 projects showed that no single proxy was found to be a good measure for stage effort prediction. This result emphasizes the need to make appropriate decisions regarding proxy selection. The authors came to conclude that it is difficult to improve stage effort prediction during software development, at least if the early estimation was fairly good.

In the opposite direction, Wang and colleagues [12] built a gray learning method based on GM(1,1) for stage effort prediction, where the stage in their study was a calendar month. They claimed it is more frequently used by managers to plan and control the progress of a project. Thus, the manager needs to predict project schedule regularly. Results on 10 datasets demonstrated that the model has a considerable potential to deliver good stage estimation.

## 3. FUZZY SET THEORY

Fuzzy set theory as introduced by Zadeh [25] provides a representation scheme and mathematical operations for dealing with uncertain, imprecise and vague concepts. Fuzzy logic is a combination of a set of logical expressions with Fuzzy sets. Zadeh [25] defined the meaning of the membership for Fuzzy sets to be a continuous number between zero and one. Each Fuzzy set is described by membership function such as Triangle, Trapezoidal, Gaussian, etc., which assigns a membership value between 0 and 1 for each real point on universe of discourse.

## 4. ASSOCIATION RULE MINING

Association rule mining is one of the important techniques in data mining [19] which aims to discover the associations and frequent patterns amongst set of items in a particular database [20, 21]. It has been successfully applied in various fields such as: market management [19], product purchasing logs of retail stores [20], website traffic logs [21] and classification [22]. The association rules do not imply causality which means that each rule is attached with a weight that relates to the statistical confidence of this rule. Association rule is denoted by an expression (A=>B) where $A$ is Antecedent and $B$ is Consequent, both $A$ and $B$ are sets of items [20]. For example, in an online book store there are always some tips displayed when you purchase a particular book containing a list of some related books as recommendation for further purchasing. Below we explain the association rule technique in more details:

Let $D$ be database of different transaction records, $I= \{I1, I2, I3, I4…, Im\}$ be a set of $m$ distinct binary attribute values called items [19, 20]. Each transaction $T \in D$ is a set of items such that $T \subseteq I$. Association rule is an implication in the form $A=>B$ which means that whenever $T$ contains $A$, then $T$ also contains $B$ with specified confidence [21, 22], where $A, B \subset I$ are sets of items called itemsets. Since the data base is large and users only concern about frequent interesting patterns, there are two measures used to capture the statistical strength of a pattern: support and confidence [19, 20]. Support is an indicator of rule frequency. The rule confidence is the probability that consequent $B$ will follow antecedent $A$ and is expressed as the percentage of transactions containing $A$ and $B$ to the overall number of transactions containing $A$. The pre-defined thresholds for interesting association rule are called minimal support and minimal confidence respectively [19, 20, 21].

Most of association rule mining algorithms are not applicable to software engineering data because these data are often represented in numeric scale but the algorithms deal only with categorical (nominal) data [20, 23]. In this paper we would like to extend the association rule technique to take of advantage of the numeric values by distributing them to intervals as discussed in the next section, and then represent each interval with nominal data. All extracted association rules should be filtered according to target stage. For example, if the target stage prediction is the "*design phase*" then we have to filter all extracted rules that contain design phase as consequent only.

## 5. THE PROPOSED APPROACH

The proposed approach combines the concepts of Fuzzy set theory [25] and association rule mining [19, 20]. The Fuzzy set theory [25] is used to represent the corresponding linguistic variables for each interval instead of representing them as crisp interval. Thus, this should help us to derive the final prediction after determining the corresponding Fuzzy set for the target stage. Determining the corresponding Fuzzy set for the target stage is performed by using association rule which attempts to find confident rules between prior stage(s) and stage under prediction. The approach is described by 5 steps as explained below:

**Step1**: define the universe of discourse U for each stage in historical dataset, then divide it into several equal intervals (lengths). In this step the minimum ($D_{min}$) and maximum ($D_{max}$) value of each universe of discourse is determined. Consequentially, based on $D_{min}$ and $D_{max}$ we define the universe U as $[D_{min}-D_1, D_{max}+D_2]$ where $D_1$ and $D_2$ are two proper positive numbers used to make the universe U more clear containing all possible values in the dataset. After that, each U should be partitioned into a number of equal intervals where the number and length of intervals should be predefined by estimator. Assuming $n$

is the number of intervals then the length of interval $L$ is calculated as follows:

$$L = \frac{[(D_{\max} + D_2) - (D_{\min} - D_1)]}{n} \quad (1)$$

Then each interval is defined as follows:

$$W_i = [(D_{\min} - D_1) + (i-1)L, (D_{\max} - D_2) + iL], i \leq n \quad (2)$$

For example, let us consider the "*specification stage*" has the following boundary: ($D_{min}$=22) and ($D_{max}$=162). For simplicity we choose $D_1$=12 and $D_2$=8, thus the universe of discourse for the specification stage is defined as U=[10, 170]. This means that based on available historical data the effort records of specification stage is delimited between 10 to 170 man-months. Let U be divided into four equal intervals with equal length as following:

$$L = \frac{[(162+8) - (22-12)]}{4} = 40$$

with: $W_1$ =[10, 50), $W_2$ =[50, 90), $W_3$ =[90, 130) and $W_4$ =[130, 170).

**Step 2**: define a corresponding linguistic variable (Fuzzy set) for each interval in the universe of discourse U. The number of Fuzzy sets must be related to the number of intervals. Let $A_1, A_2, A_3, \ldots, A_n$ be Fuzzy sets which are linguistic terms defined as depicted in equation 3:

$$A_i = \{(\mu_{A_i}(W_j)/W_j) | \mu_{A_i}(W_j) \in [0,1], W_j \in R, 1 \leq j \leq n, 1 \leq i \leq n\} \quad (3)$$

where:
- $\mu_{A_i}(W_j)$ is the membership degree of interval $W_j$ in Fuzzy set $A_i$.
- $n$ corresponds to the number of intervals.

Therefore the linguistic terms $A_1, A_2, A_3, \ldots, A_n$ will be defined as follows:

$A_1 = \{1/W_1, 0.5/W_2, 0/W_3, 0/W_4, \ldots 0/W_{n-2}, 0/W_{n-1}, 0/W_n\}$

$A_2 = \{0.5/W_1, 1/W_2, 0.5/W_3, 0/W_4, \ldots 0/W_{n-2}, 0/W_{n-1}, 0/W_n\}$

… … … …
… … … …

$A_n = \{0/W_1, 0./W_2, 0/W_3, ./0/W_4 \ldots 0/W_{n-2}, 0.5/W_{n-1}, 1/W_n\}$

Based on the previous example in step 1, the possible Fuzzy sets for the four intervals $W_1, W_2, W_3, W_4$ should be defined as follows:

$A_1 = \{1/W_1, 0.5/W_2, 0/W_3, 0/W_4\}$

$A_2 = \{0.5/W_1, 1/W_2, 0.5/W_3, 0/W_4\}$

$A_3 = \{0/W_1, 0.5/W_2, 1/W_3, 0.5/W_4\}$

$A_4 = \{0/W_1, 0/W_2, 0.5/W_3, 1/W_4\}$

**Step 3**: determine the target stage and discover association rules between prior stage(s) and target stage. In this step we used predictive APRIORI algorithm [22] that is implemented in WEKA data mining tool [26]. The minimum support is set by **0.01** and minimum confidence is set by **0.8**. These values have been carefully chosen to avoid too few rules that would occur if the confidence was very high.

In this paper we will replace the name of all stages with the following abbreviations. The number preceding the abbreviation represents the order of stage in software development process.

1. **EP**: Effort of Planning stage.
2. **ES**: Effort of Specification stage.
3. **ED**: Effort of Design stage.
4. **EB**: Effort of Building stage.
5. **ET**: Effort of Testing stage.
6. **EI**: Effort of implementation stage.

**Step 4**: filtering extracted rules. All generated rules are filtered to obtain interesting rules that contain specified target as consequent and all rules should respect stage order integrity. This means that all stages in antecedent parts should not precede target stage in consequent part. For example, if the target stage is the "*design phase: ED*" then all rules that contain this phase only will be taken for further processing and others are neglected. The following rules are taken for further processing:

EP1=>ED4

ES2 and EP3=>ED2

The number after abbreviation denotes corresponding Fuzzy set (interval).

Conversely, the following rules are neglected because there are problems in either antecedent or consequent part:

EP1 and ES2=>ED1 & ET3: because ED1 should appear alone in consequent part

ES1 and EI=>ED1: because EI cannot precede ED

**Step 5**: calculate the predicted output. Firstly, defuzzify all expected outputs with regards to target stage:

$$defuzz(A_i) = \frac{\sum_{j=1}^{n} \mu_{A_i}(W_j) * m(W_j)}{\sum_{j=1}^{n} \mu_{A_i}(W_j)}, \forall i = \{1,2\ldots n\} \quad (4)$$

where $m(W_j)$ is the centre value of expected interval of target stage in historical dataset. Secondly, the estimated effort is calculated by computing the weight average of defuzzification values. The weight here is confidence ratio of extracted rules as shown in equation 5.

$$\hat{E} = \frac{\sum_{i=1}^{k} defuzz(A_i) * rule\ confidence_i}{\sum_{i=1}^{k} rule\ confidence_i} \quad (3)$$

where *k* is the number of rules.

For example assume we want to predict specification stage of a project. Consider prior stage is software plan phase and its effort value is located in the first interval (EP1). Based on association rule, the following rules have been extracted:

EP1=>ES4 (confidence= 0.932)

EP1=>ES3 (confidence= 0.843)

EP1=>ES1 (confidence= 0.78)

Then corresponding Fuzzy sets that represent expected target stage based on previous rules should be defuzzfied. From this example we can observe that the input interval has many relations with target intervals, i.e. EP1 has three significant relations with ES4, ES3 and ES2 in the specification phase. Therefore we need to take their impacts on the final estimate. The effort for specification phase stage is calculated as following:

$$defuzz(A_{ES4}) = \frac{0*m(W_1)+0*m(W_2)+0.5*m(W_3)+1*m(W_4)}{0+0+0.5+1}$$

$$defuzz(A_{ES3}) = \frac{0*m(W_1)+0.5*m(W_2)+1*m(W_3)+0.5*m(W_4)}{0+0.5+1+0.5}$$

$$defuzz(A_{ES1}) = \frac{1*m(W_1)+0.5*m(W_2)+0*m(W_3)+0*m(W_4)}{1+0.5+0+0}$$

Assume $m(W_1)=20, m(W_2)=35, m(W_3)=55, m(W_4)=70$ then

$defuzz(A_{ES4}) = 65$.

$defuzz(A_{ES3}) = 53.75$

$defuzz(A_{ES1}) = 25$.

By using equation 5 the predicted effort is:

$$\hat{E} = \frac{0.932*defuzz(A_{ES4})+0.843*defuzz(A_{ES3})+0.78*defuzz(A_{ES2})}{0.932+0.843+0.78}$$

$\hat{E} = 49.08 \quad man-months$

## 6. EVALUATION CRITERIA

Many evaluation criteria are introduced in software engineering literature, among them we selected three evaluation criteria are Bias, Mean Magnitude of relative errors (MMRE) and Median Magnitude of relative errors (MdMRE). Bias in equation (6) is used to check whether the proposed prediction model is biased and tends to under or over estimation. MMRE in equation (7) computes the degree of estimation error in an individual estimate and should be less than 25% to be acceptable. Since the MMRE is sensitive to the individual prediction with large MRE we adopt median MRE (MdMRE) which is less sensitive to the extreme value of MRE. The acceptable target for MMRE and MdMRE is less or equal to 25%.

$$Bias(i) = \frac{actual_i - estimated_i}{actual_i} \quad (4)$$

$$MMRE = \frac{1}{n}\sum_{i=1}^{n}|Bias(i)| \quad (6)$$

$$MdMRE = median(|Bias(i)|), \forall i \quad (5)$$

## 7. RESULTS AND DISCUSSIONS

The dataset used in empirical validation came from ISBSG [24]. The obtained dataset contains effort records for six phases are: plan effort, specification effort, design effort, building effort, testing effort, implementation effort. As a preliminary stage of data pre-processing we attempted to select the most representative data, therefore we ignored the projects records that contain missing values.

Determining the possible number of intervals in each stage is carried out based on the distribution of effort data in each stage as shown in Figures 1 to 6. There is no clear mechanism for how to determine the perfect number of intervals therefore we attempted to study density of data for each stage separately. The performed analysis resulted in different number of intervals between stages. The obtained number of intervals reflects the density and range of data in each stage.

**Table 1.** Number of intervals

| Stage | Number of Intervals |
|---|---|
| Planning | 7 |
| Specification | 8 |
| Design | 10 |
| Building | 9 |
| Testing | 8 |
| Implementation | 11 |

The theme of this paper is to address the following arising issue: can project manager relay on prior effort records to predict next stage effort? To answer this question, the proposed model has been evaluated using jack-knifing method. We used 34 projects with complete effort records.

Table 2 and Table 3 depict the results obtained by our proposed approach compared to exponential regression (where target stage is regarded as dependent variable and all pervious stages as independent variables). From Table 2 we can observe that all outputs tend to be under estimation. Three out of five stages producing good estimate are specification, building and testing, while design stage produced better results compared to implementation stage (which produced the worst stage effort estimation in terms of MMRE). The reason is related to that the ISBSG is scattered as result of collection from different worldwide companies. The effort records have complex structure in which there is no consistent structure for all effort records. Based on MdMRE we can observe that our approach in most of stages produced comparable estimation accuracy with maximum 30.2% in implementation stage. Results shown in Table 3 revealed that most of predictions are under estimation which supports our approach findings. The best estimation accuracy was obtained in building stage, which also corroborates our findings that best estimation accuracy was in building stage. The negative values in Bias criterion show underestimation. It is acknowledged that MMRE is unbalanced in many validation circumstances and leads to overestimation more than underestimation. In our case, we found that MMRE leads to underestimation in most stages. This is may be related to the absence of systematic scheme between all prior effort records.

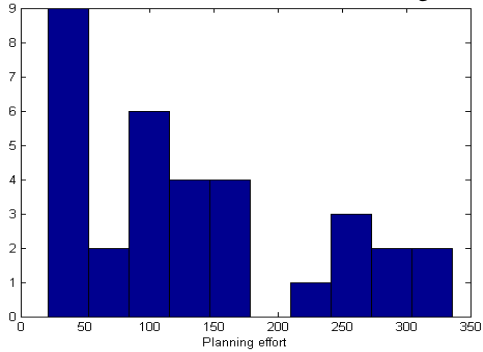

**Fig. 1.** Effort distribution of Planning stage

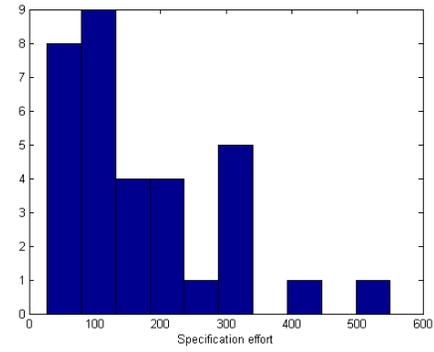

**Fig. 2.** Effort distribution of Specification stage

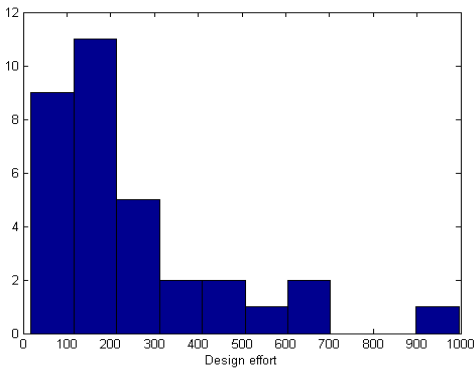

**Fig. 3.** Effort distribution of Design effort stage

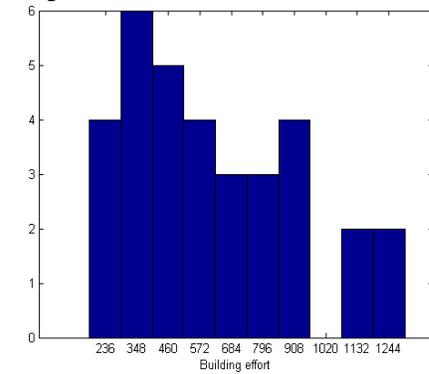

**Fig. 4.** Effort distribution of Building stage

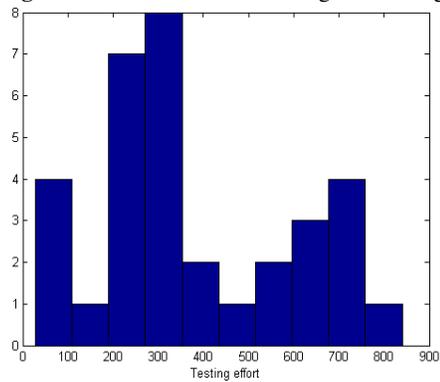

**Fig. 5.** Effort distribution of Testing stage

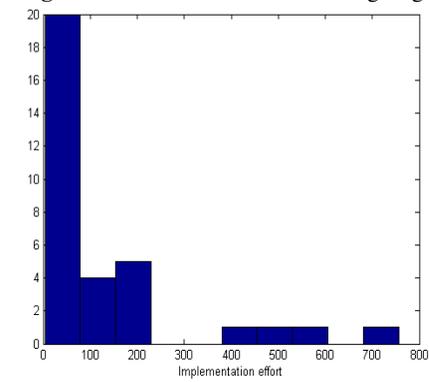

**Fig.6.** Effort distribution of Imp. stage

**Table 2.** Results using the proposed approach

| Stage | Bias | MMRE | MdMRE |
|---|---|---|---|
| Specification effort | -8.5% | 27.0% | 17.0% |
| Design effort | -33.1% | 40.5% | 13.7% |
| Building effort | -2.8% | 9.3% | 7.5% |
| Testing Effort | -11.6% | 16.7% | 7.23% |
| Implementation effort | -20% | 91.0% | 30.2% |

**Table 3.** Results using exponential regression

| Stage | Bias | MMRE | MdMRE |
|---|---|---|---|
| Specification effort | -24.3% | 81.3% | 49.7% |
| Design effort | -72.3% | 120.4% | 54.224% |
| Building effort | 0.7% | 44.35% | 37.6% |
| Testing Effort | -45.4% | 81.1% | 39% |
| Implementation effort | -179% | 184% | 104% |

The comparison between our approach and exponential regression technique showed that there are considerable improvements in estimation accuracy on all phases of software development lifecycle. MMREs of our approach have been reduced by at least 35.05% and at most 93%. Biases have been reduced by at least 3.5% and at most 159%. We have to bear in mind that the length of interval plays important role in estimation accuracy, thus, when the universe of discourse is partitioned into several equal intervals, the distribution of data should be taken into account. Moreover, we should remove the extreme values because they affect interval partitioning, thus, estimation accuracy.

Figures 7 to 11 show comparison between proposed approach and exponential regression in each stage by using Boxplot. The Boxplot [17] offers a way to compare between estimation models based on their absolute residuals. The Boxplot is non-parametric statistics used to show the median as central tendency of distribution, interquartile range and the outliers of individual models [17]. The length of Boxplot from lower tail to upper tail shows the spread of the distribution. The length of box represents the interquartile range that contains 50% of observations. The position of median inside the box and length of Boxplot indicate the skewness of distribution. A Boxplot with a small box and long tails represents a very peaked distribution while a Boxplot with long box represents a flatter distribution.

The prominent and common characteristic among these figures is the spread of absolute residuals for our approach is less than spread of exponential regression which presents more accurate results. The larger interquartile of exponential regression indicates a high dispersion of the absolute residuals. The Boxplot revealed that the box length for our models is smaller than exponential regression which also indicates reduced variability of absolute residuals. The median of our model is smaller than median of exponential regression which revealed that at least half of the predictions of our model are more accurate than exponential regression. The lower tails of our model is much smaller than upper tail which means the absolute residuals are skewed towards the smaller value.

Figure 11 illustrates the reason of why prediction of implementation stage in our approach produced the worst accuracy. The reason related to the existing of outlier. Although one project is considered as an outlier the MMRE is easily influenced with that project.

Based on the obtained results, we can observe that exponential regression gave bad accuracy. The reason may relate to the structure complexity of prior effort records. There is no correlation between all prior stages and target stage.

To ensure that the results obtained are not by chance we investigated the statistical significance of the proposed approach using Wilcoxon sum rank test for absolute residuals as shown in Table 4. In this test if the resulting $p$-value is small ($p<0.05$), then a statistically significant difference can be accepted between the two samples' median. The residuals obtained using the proposed approach were significantly different from those obtained by exponential regression. Suggesting that, there is difference if the predications generated using the proposed approach or exponential regression and based on the accuracy comparison in Tables 2 and 3 we can safely conclude that our proposed method outperformed exponential regression for stage effort estimation.

**Table 4.** Statistical significance

| Stage | sum rank | Z-value | p-Value |
|---|---|---|---|
| Specification effort | 769 | -4.31 | <0.01 |
| Design effort | 713 | -5.03 | <0.01 |
| Building effort | 685 | -5.4 | <0.01 |
| Testing Effort | 595 | -6.54 | <0.01 |
| Implementation effort | 799 | -3.93 | <0.01 |

As in any experiment, there always some of threats affect empirical validation. In our case:

1. the proposed model is validated only over ISBSG data, thus we believe is not sufficient. There is need for more investigation based on data collected specially for stage effort estimation purpose.
2. the major threat to validity of our study is the population model. It is very hard to choose representative data; we performed pre-processing stage to identify the most representative data by ignoring projects that contain missing values in all effort records. It is argued that removing those projects could loss some valuable information.
3. length of interval and existing of outliers. The extreme values has significant impact of intervals partitioning therefore it leads to bad estimation accuracy. Most of extreme values in all universe of discourse have been removed which resulted in 34 representative projects.
4. number of rules: when number of prior stages increase, the number of extracted rules will be also increased. Furthermore, sometimes the number of rules is too few because of minimum support and confidence. Thus it becomes difficult to predict the target stage effort unless we change minimum confidence.

## 8. CONCLUSIONS

Some of software projects are failed due to the absence of re-estimation during software development which results in huge gap between initial plan and final outcome. Even with good estimate at first stage the project manager must keep update with project progress and should be able to re-estimate the project at any particular point of project in order to re-allocate the proper number of resources. The objective of this paper was to check whether the prior effort records can be used to predict stage effort with reasonable accuracy or not. The obtained results revealed that using association rule and Fuzzy set theory lead to significant improvement in stage-effort estimation and give project manager an evolving picture about project progress. Comparing our approach with exponential regression showed that there is a considerable potential in estimation accuracy. As part of future plan, we intend to expand this work to involve some interesting features in each stage prediction and evaluate it on many datasets.

## 9. ACKNOWLEDGMENTS

Authors would like to thank ISBSG for granting us permission to use their dataset.

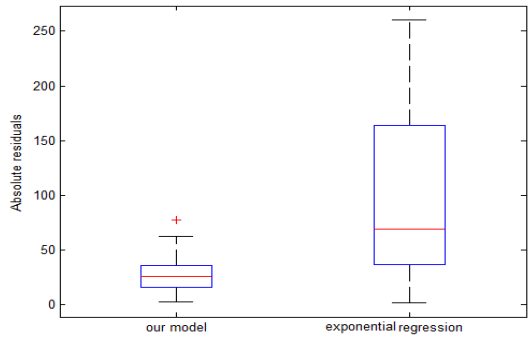

**Fig. 7.** Boxplot of absolute residuals for the specification stage

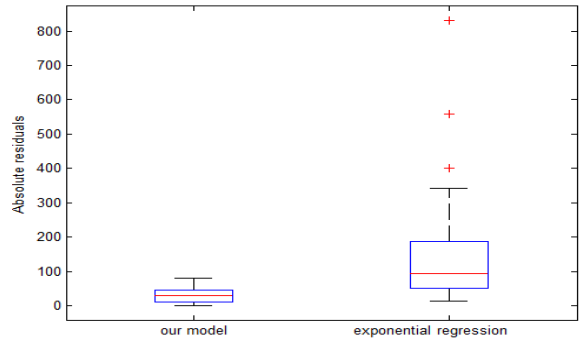

**Fig. 8.** Boxplot of absolute residuals for the design stage

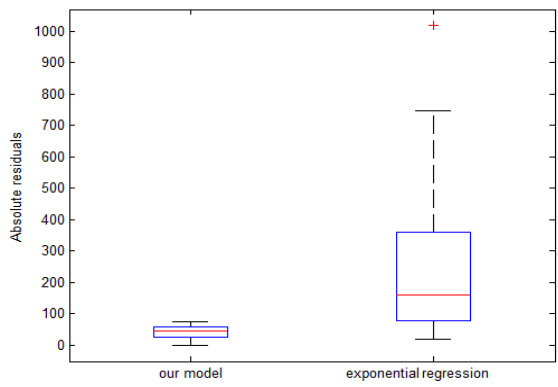

**Fig. 9.** Boxplot of absolute residuals for the building stage

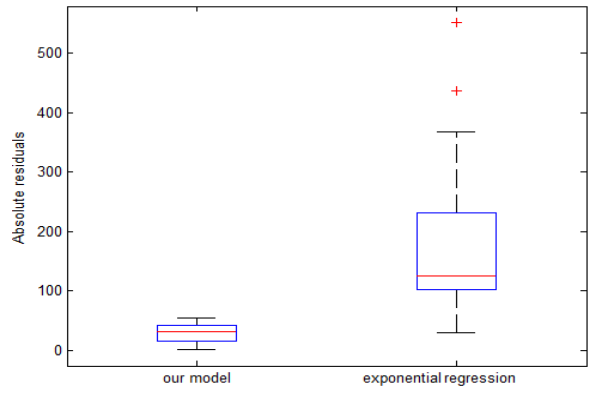

**Fig. 10.** Boxplot of absolute residuals for the testing stage

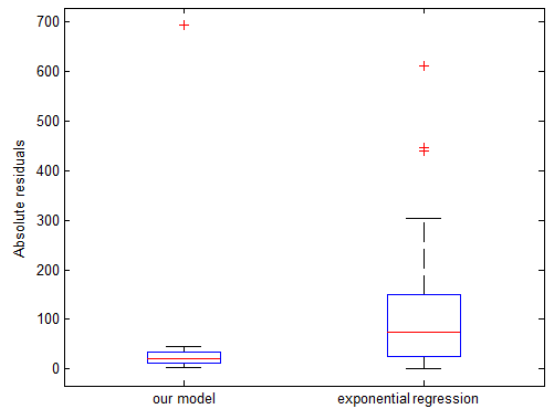

**Fig. 11.** Boxplot of absolute residuals for the implementation stage